\documentstyle[12pt]{article}
\newlength{\aivwidth}   
\newlength{\tmpwidth}   

\newcommand{\phr}[1]{Phys.\ Rev.\ {\bf #1}}
\newcommand{\phrd}[1]{Phys.\ Rev.\ {\bf D#1}}
\newcommand{\phrl}[1]{Phys.\ Rev.\ Lett.\ {\bf #1}}
\newcommand{\nphb}[1]{Nucl.\ Phys.\ {\bf B#1}}
\newcommand{\phlb}[1]{Phys.\ Lett.\ {\bf B#1}}
\newcommand{\zphc}[1]{Z.\ Phys.\ {\bf C#1}}
\newcommand{\aph}[1]{Ann.\ Phys.\ {\bf #1}}
\newcommand{\phrp}[1]{Phys.\ Rep.\ {\bf #1}}
\newcommand{\ptph}[1]{Prog.\ Theor.\ Phys.\ {\bf #1}}

\newcommand{\be}{\begin{equation}}
\newcommand{\ee}{\end{equation}}
\newcommand{\bea}{\begin{eqnarray}}
\newcommand{\eea}{\end{eqnarray}}
\newcommand{\ba}{\begin{array}}
\newcommand{\ea}{\end{array}}
\newcommand{\eref}[1]{(\ref{#1})}
\newcommand{\nn}{\nonumber\\}

\newcommand{\eqnew}{\setcounter{equation}{0}}

\newcommand{\dfun}{{\cal D}}
\newcommand{\lag}{{\cal L}}

\newcommand{\dx}{d^4x\,}
\newcommand{\ep}{\epsilon}
\newcommand{\vp}{\varphi}
\newcommand{\pa}{\partial}

\title{Effective Lagrangians\\with Higher
Order Derivatives}
\author{Carsten Grosse-Knetter\thanks{E-mail:
knetter@physw.uni-bielefeld.de}\\[5mm] Universit\"at
Bielefeld\\Fakult\"at f\"ur Physik\\33501 Bielefeld\\Germany}
\date{BI-TP 93/29\\hep-ph/9306321\\June 1993}

\begin{document}
\begin{titlepage}
\maketitle
\thispagestyle{empty}
\begin{abstract}
The problems that are connected with Lagrangians which depend on higher order
derivatives (namely additional degrees of freedom, unbound energy from below,
etc.)\ are absent if {\em effective\/} Lagrangians are considered because
the equations of motion may be used to eliminate all higher order time
derivatives from the effective interaction term. The application of the
equations of motion can be realized by performing field transformations that
involve derivatives of the fields. Using the Hamiltonian formalism for higher
order Lagrangians (Ostrogradsky formalism), Lagrangians that are related by
such transformations are shown to be physically equivalent (at the classical
and at the quantum level). The equivalence of Hamiltonian and Lagrangian path
integral quantization (Matthews's theorem) is proven for effective higher
order Lagrangians. Effective interactions of massive vector fields involving
higher order derivatives are examined within gauge noninvariant models as well
as within (linearly or nonlinearly realized) spontaneously broken gauge
theories. The Stueckelberg formalism, which relates gauge noninvariant to gauge
invariant Lagrangians, becomes reformulated within the Ostrogradsky formalism.
\end{abstract}
\end{titlepage}


\section{Introduction}
\typeout{Section 1}
\eqnew
Effective Lagrangians containing arbitrary interactions of massive
vector and scalar fields are studied very intensively in the
literature in order to
parametrize possible deviations of electroweak
interactions from the standard model with respect to experimental
tests of the $W^\pm$, $Z$ and $\gamma$ self-interactions and of the
Higgs sector. When performing a complete analysis of
the extensions of
the standard (Yang--Mills) vector-boson self-interactions
to nonstandard interactions,
one necessarily has  to consider effective interaction
terms that depend on higher order derivatives of the fields
\cite{apbe,phen}.
Therefore, the investigation of effective
Lagrangians with higher order derivatives is very important from the
phenomenological point of view.

However, theories described by
higher order Lagrangians have quite unsatisfactory properties
\cite{ost,hide,bedu}, namely: there are
additional degrees of freedom, the energy is unbound from below,
the solutions of the equations of motion are not uniquely determined
by the
initial values of the fields and their first time derivatives and
the theory has no analytic limit for $\epsilon\to 0$
(where $\epsilon$ denotes the coupling constant
of the higher order term).
Clearly, these features are very unwelcome
when dealing with effective Lagrangians in order to parametrize {\em
small\/} deviations from a renormalizable theory like the standard
model.

Fortunately however,
the abovementioned problems are absent if a
higher order Lagrangian is considered to be an {\em effective\/}
one. This means, one assumes that there
exists a renormalizable theory with
heavy particles at an energy scale $\Lambda$
(``new physics''), and that the effective Lagrangian
parametrizes the effects of the ``new physics'' at an energy
scale lower than $\Lambda$ by expressing the contributions
of the heavy fields (which do not explicitely occur in the effective
Lagrangian) through nonrenormalizable effective interaction
terms. Supposed that the renormalizable Lagrangian
describing the ``new physics'' does
not depend on higher order derivatives,
it causes no unphysical effects and
therefore such effects also do not occur at the lower energy scale,
i.e.\ at the effective Lagragian level.
Actually, I will show in this paper that
all higher order time derivatives can be eliminated
in the first order of the effective coupling constant
$\epsilon$ (with $\epsilon\ll 1$).
Higher powers of $\epsilon$ can be neglected
because an effective Lagrangian
is assumed to describe the effects of  well-behaved ``new physics''
the $O(\ep)$ approximation only; consequently
all ill-behaved effects (which do
not occur in the first order of $\epsilon$) become cancelled by
other $O(\epsilon^n)$ $(n>1)$ effects of the ``new physics''.

Each effective higher
order Lagrangian can be reduced to a first order Lagrangian because
one can apply the equations of motion (EOM) in order to
eliminate all higher order time derivatives from the effective
interaction term (upon neglecting higher powers of $\epsilon$).
This is a nontrivial
statement because, in general, the EOM
must not be used to convert the Lagrangian. However, it has been
shown that it is possible to find
field transformations which have the
the same effect as the application
of the EOM to the effective interaction term
(in the first order of
$\epsilon$) \cite{eom,pol,geo}.
Within the Hamiltonian formalism for higher order Lagrangians
(Ostrogradsky formalism) \cite{ost}, these field transformations
are point transformations (and thus canonical
transformations) although they involve derivatives of the fields.
The reason for this is that, within the Ostrogradsky formalism,
the {\em derivatives\/} up to order $N-1$
are formally treated as independent {\em coordinates\/} if the
Lagrangian is of order $N$, and the order
$N$ of the Lagrangian can be chosen arbitrarily
 without affecting the physical content of the theory
\cite{gity} (as long as $N$ is
greater or equal to the order of the highest actually appearing
derivative).  This implies that the Lagrangian which is obtained
from the primordial Lagrangian by applying the EOM
to its effective interaction term is
indeed physically equivalent to this (at the classical and
at the quantum level).
In this paper I will describe this method of reducing effective
higher order Lagrangians.

The main task of this paper is the quantization of effective
higher order Lagrangians within the Feynman path integral (PI)
formalism. It is well known that, in general, quantization has to be
based on the Hamiltonian PI \cite{gity,ham}
and not on the naive Lagrangian PI.
The Lagrangian PI, however, is more useful for practical
calculations because it does not involve the momenta and it
is manifestly covariant.
In fact, it has been shown for
arbitrary effective interactions of scalar fields \cite{bedu} and of
massive vector fields\footnote{Using the same procedure as in
\cite{gk} this result can also be derived
for the (phenomenologically less interesting) case of effective
interactions involving fermion fields.}
\cite{gk} (which depend on first order
derivatives of the fields only)
that, after the momenta have been integrated out,
the Hamiltonian PI corresponding to an effective Lagrangian $\lag$
can be written as simple Lagrangian PI
\be
Z=\int\dfun\varphi\,\exp\left\{i\int\dx[\lag_{quant}+J\varphi
]\right\}
\label{lpi}\ee
($\varphi$ is a shorthand notation for all fields in
$\lag_{quant}$), where
the quantized
Lagrangian $\lag_{quant}$ is identical to the primordial
Lagrangian $\lag$:
\be
\lag_{quant}=\lag
\label{eq}\ee
(if $\lag$ has no gauge freedom).
This means that the correct Hamiltonian PI
quantization yields the same
result as naive Lagrangian PI quantization. Actually,
this result justifies
Lagrangian PI quantization. Furthermore, (\ref{lpi}) and (\ref{eq})
imply that the Feynman rules follow
directly from the effective Lagrangian, i.e., the quadratic
terms in $\lag$ yield the propagators and the other terms yield the
vertices in the standard manner. This simple quantization
rule is known as Matthews's theorem \cite{mat}. Within
the PI formalism Matthews's theorem
can be reformulated as the statement
that Hamiltonian and Lagrangian PI quantization
are equivalent.

Matthews's theorem has yet
only been proven for effective
interactions which involve at most first order derivatives
\cite{bedu,gk}. In the higher order case, the result
\eref{lpi}, \eref{eq} has been derived for some special
examples \cite{bedu,bnn} but not as a general theorem.
In this paper I will prove Matthews's theorem for
Lagrangians with arbitrary
higher order effective interactions making use of
the abovementioned procedure of reducing higher order effective
Lagrangians to first order ones and then applying Matthews's theorem
for Lagrangians without higher order derivatives.

I will derive this result first for the simple case of effective
interactions of scalar fields and then extend it to effective
interactions of massive vector fields. In the latter case one has to
distinguish between gauge noninvariant effective Lagrangians  and
(nonlinearly or linearly realized) spontaneously broken gauge
theories (SBGTs). However, the results obtained for gauge noninvariant
Lagrangians can easily be extended to gauge invariant Lagrangians
because, by applying a Stueckelberg transformation
\cite{stue,bulo,gkko}, each SBGT can be rewritten as a gauge
noninvariant model (after a nonlinear parametrization of the
scalar sector \cite{gkko,lezj,clt}
in the case of a linear Higgs model).
A Stueckelberg transformation
is a field transformation (that involves
derivatives of the fields) which results in removing all
unphysical scalar fields from the Lagrangian. In \cite{gk}, I have
proven the canonical
equivalence of Lagrangians (without higher order
derivatives) that are related by a Stueckelberg transformation.
Within the Ostrogradsky formalism this result can be
generalized to higher order Lagrangians and besides the
proof can be simplified very much
since, as mentioned above, within this formalism a field
transformation which involves derivatives is a
point transformation, i.e. canonical transformation.
For gauge invariant models, Matthews's theorem states
that an effective theory can be quantized using the
(Lagrangian) Faddeev--Popov formalism \cite{fapo}. The quantized
Lagrangian in \eref{lpi} becomes
\be
\lag_{quant}=\lag+\lag_{g.f.}+\lag_{FP-ghost},
\label{fapo}\ee
i.e., there are additional gauge fixing (g.f.)\
and ghost terms. This result can be derived as follows:
using the Stueckelberg formalism and
Matthews's theorem for gauge noninvariant Lagrangians
one can show that the Hamiltonian
PI corresponding to a SBGT is identical to
the one obtained within the
Faddeev--Popov formalism, if
the (U-gauge) g.f. conditions that all
unphysical scalar fields become equal to zero are imposed%
\footnote{For the special case of the U-gauge, the
Faddeev--Popov procedure does not yield explicit g.f.\ and
ghost terms \cite{gkko} unlike in the general case \eref{fapo}.}
\cite{gkko}.
The equivalence of all gauges,
i.e.\ the independence of the $S$-matrix
elements from the choice of the gauge in the Faddeev--Popov
procedure \cite{lezj,able}, yields then the result
\eref{lpi} with \eref{fapo} in an arbitrary gauge.

As in the case of Lagrangians with at most first order derivatives,
the result \eref{eq} or \eref{fapo} is only correct up to additional
terms
proportional to $\delta^4(0)$ \cite{bedu,gk} which, however, become
zero if dimensional regularization is applied. According to
\cite{bedu,gk}, I will neglect all these $\delta^4(0)$ terms
in this paper.

This paper is organized as follows: In section~2,  the
Hamiltonian formalism for Lagrangians with higher order
derivatives (Ostrogradsky formalism) and the Hamiltonian path
integral quantization of such Lagrangians are briefly reviewed.
Using this formalism,
Lagrangians which are related by
field transformations that involve derivatives
of the fields are shown to be equivalent
(at the classical and at the quantum level).
In section~3, it is shown that
higher order effective Lagrangians
can be reduced to first order
ones by applying the equations of motion.
The equivalence of Hamiltonian and Lagrangian path integral
quantization (Matthews's theorem)
is proven for higher order effective interactions
of scalar fields.
These results become extended to
the case of massive vector fields in section~4.
Spontaneously broken gauge theories
are reduced to gauge noninvariant Lagrangians using the
Stueckelberg formalism, which becomes reformulated within the
Ostrogradsky formalism.
Section~5 contains the final conclusions.


\section{The Ostrogradsky Formalism}
\typeout{Section 2}
\eqnew
The Hamiltonian formalism for Lagrangians with higher
order derivatives has been formulated one and a half century ago by
Ostrogradsky \cite{ost}. I will briefly review this formalism, since
it is not as well known as the Hamiltonian
formalism for Lagrangians with at most first order derivatives.
I will also take into account the case of
singular higher order Lagrangians \cite{gity,shd}, i.e.\ of
Lagrangians which involve constraints\footnote{The reader
who is not familiar with the basic concepts of dynamics
and quantization of constrained systems is refered to
\cite{gity,ham,dirac}.}.
Furthermore, I will consider the quantization of higher order
Lagrangians within the Hamiltonian PI formalism.
For simplification I will restrict to the case of
finitely many degrees of freedom; the generalization to field theory
is analogous to the first order case.

Consider a Lagrangian of order $N$, i.e.\ a Lagrangian which depends
on the coordinates $q_i$, with $i=1,\ldots,I$ and their
derivatives up to the order $N$:
\be
L(q_i,q_i^{(1)},\ldots,q_i^{(N)}),\qquad\mbox{with}\qquad
q_i^{(n)}\equiv
\left(\frac{d}{dt}\right)^n q_i.
\label{hol}\ee
Within in the Hamiltonian formalism, the coordinates are
\be
Q_{i,n}\equiv q_i^{(n-1)},\qquad n=1,\ldots,N
\label{Q}\ee
and the momenta are
\be
P_{i,n}\equiv \sum_{k=n}^N\left(-\frac{d}{dt}\right)^{k-n}
\frac{\partial L}{\partial q_i^{(k)}},\qquad
n=1,\ldots,N.
\label{P}\ee
The Hamiltonian is given by
\be
H\equiv\sum_{i=1}^I\sum_{n=1}^NP_{i,n}\dot{Q}_{i,n}-L
=\sum_{i=1}^I
\left(\sum_{n=1}^{N-1}P_{i,n}Q_{i,n+1}+P_{i,N}q_i^{(N)}
\right)-L.
\label{ham}
\ee
In \eref{ham}, the $q_i^{(N)}$
have to be expressed in terms of the
$P_{i,N}$ by using \eref{P} with $n=N$:
\be
P_{i,N}=\frac{\partial L}{\partial q_i^{(N)}}.
\label{PN}\ee
If the theory is nonsingular, i.e. if the $I\times I$-matrix
\be
M_{ij}\equiv\frac{\partial L}{\partial q_i^{(N)}
\partial q_j^{(N)}}
\label{sing}\ee
is nonsingular, \eref{PN} can be solved for all $q_i^{(N)}$. For
singular Lagrangians one finds
\be
\mbox{rank}\,M_{ij}=R<I.
\label{R}\ee
The indices $i,j$
can be ordered such that the upper left
$R\times R$-submatrix
of $M$ has the rank $R$. In this case, \eref{PN} can be
solved for $q_i^{(N)}$
with $i=1,\ldots,R$ only, while the remaining of these
relations are constraints.
However, $H$ \eref{ham} does
not depend on $q_i^{(N)}$ with $i=R+1,\ldots,I$ because
\be
\frac{\partial H}{\partial q_i^{(N)}}=
P_{i,N}-\frac{\partial L}{\partial q_i^{(N)}}=0,
\qquad i=R+1,\ldots,I\, ,
\ee
where \eref{PN} has been used.
Furthermore, \eref{R} implies that the first $R$ of the relations
\eref{PN} can be used to rewrite the remaining ones in a form which
does not involve the $q_i^{(N)}$.
The total Hamiltonian $H_T$ is defined as
\be
H_T\equiv H+\lambda_a\phi_a,
\label{ht}\ee
where the $\lambda_a$ are Lagrange multipliers
and the $\phi_a$ are the primary constraints \eref{PN}
with $i=R+1,\ldots ,I$ and the secondary, tertiary, etc.\
constraints (which follow
from the primary constraints due to
the requirement that the constraints have to
be consistent with the EOM).
The Hamiltonian EOM can be written as
\bea
\dot{Q}_{i,n}&=&\frac{\partial H_T}{\partial P_{i,n}}
=\{Q_{i,n},H_T\},\label{qeom}\\
\dot{P}_{i,n}&=&-\frac{\partial H_T}{\partial Q_{i,n}}
=\{P_{i,n},H_T\}\label{peom}.
\eea
It should be emphasized that, within the Ostrogradsky formalism
for a Lagrangian of order $N$, {all derivatives of the $q_i$ up to
the order $N-1$ are formally treated as independent coordinates\/}
(due to \eref{Q}).

A higher order Lagrangian can be quantized within the
Hamilitonian PI formalism \cite{gity,ham}
analogously to a first order
Lagrangian. The Hamiltonian  PI turns out to be
\bea
Z&=&\int\prod_{i=1}^I\prod_{n=1}^N(\dfun Q_{i,n} \dfun P_{i,n})
\exp\left\{i\int d^4x\,\left[-H+
\sum_{i=1}^I\sum_{n=1}^N(
P_{i,n}\dot{Q}_{i,n}+
J_{i,n}Q_{i,n}+K_{i,n}P_{i,n})\right]\right\}
\nn&&\mbox{\hspace*{7cm}}\times
\prod_a\delta( \Phi_a) \,{\rm Det}^\frac{1}{2}\,\{\Phi_a,\Phi_b\}.
\label{hpi}\eea
The $J_{i,n}$ and $K_{i,n}$ are sources and the $\Phi_a$ are
the primary constraints
\eref{PN} with $i=R+1,\ldots,I$, the secondary, tertiary, etc.\
constraints and the gauge fixing
conditions if there are first class constraints.

For the following investigations it is important, that
{the Ostrogradsky formalism is not affected by changing the
(formal) order $N$ of the Lagrangian\/} as long as
$N$ is greater or
equal to the order of the highest actually appearing derivative.
In other words, the Ostrogradsky  formulations of
two physical sytems which are both given in terms of the
same Lagrangian, but one is considered to be of order $N$ and the
other to be of order $M$, are equivalent. This theorem has been
proven in \cite{gity}. I will repeat this short proof here and
I will show that also
the Hamiltonian PIs corresponding to the two
systems are identical.

It is sufficient to assume that $M=N+1$. I will now treat the
Lagrangian \eref{hol} (which does not depend on the $q_i^{(N+1)}$)
formally as an $(N+1)$st order Lagrangian. One finds the
canonical variables
\bea
\tilde{Q}_{i,n}&=& q_i^{(n-1)},\qquad n=1,\ldots,N+1,
\label{Q2}\\
\tilde{P}_{i,n}&=& \sum_{k=n}^{N+1}
\left(-\frac{d}{dt}\right)^{k-n}
\frac{\partial L}{\partial q_i^{(k)}},\qquad
n=1,\ldots,N+1.
\label{P2}\eea
With $n=1,\ldots,N$, these equations become
\be
\tilde{Q}_{i,n}=Q_{i,n},\qquad
\tilde{P}_{i,n}=P_{i,n},\qquad n=1,\ldots,N,
\label{e1}\ee
i.e., these variables are identical to the corresponding ones
obtained in the $N$th order formalism, \eref{Q} and \eref{P}.
With $n=N+1$ one finds
\bea
\tilde{Q}_{i,N+1}&=&q^{(N)}_i,\label{QN1}\\
\tilde{P}_{i,N+1}&=&\frac{\pa L}{\pa q_i^{(N+1)}}
=0.\label{pc}
\eea
The Hamiltonian is given by
\be
\tilde{H}=\sum_{i=1}^I\sum_{n=1}^{N+1}\tilde{P}_{i,n}
\dot{\tilde{Q}}_{i,n}-L
=\sum_{i=1}^I\sum_{n=1}^{N}\tilde{P}_{i,n}
\tilde{Q}_{i,n+1}-L.
\label{H2}\ee
$\tilde{H}$ is identical to $H$ \eref{ham}, except that in $H$
the $q_i^{(N)}$ are
expressed in terms of $P_{i,N}$ using \eref{PN},
but in $\tilde{H}$ they are still present
(because, due to \eref{QN1}, they are independent coordinates
in the $(N+1)$st order formalism).
The $(N+1)$st order system is singular; the relations \eref{pc}
are the primary constaints.
The requirement that these constraints have to be
consistent with the
EOM yields the secondary constraints
\be
\dot{\tilde{P}}_{i,N+1}=-\frac{\partial \tilde{H}}
{\partial \tilde{Q}_{i,N+1}}
=\frac{\partial L}{\partial q_i^{(N)}}-\tilde{P}_{i,N}=0.
\label{sc}
\ee
This is identical to the relation \eref{PN}. Applying \eref{sc}
with $i=1,\ldots,R$  in order to eliminate all
$\tilde{Q}_{i,N+1}$ from the Hamiltonian $\tilde{H}$
and from the remaining of the  constraints \eref{sc}
as outlined above\footnote{Constraints may be used to convert
the Hamiltonian and the other constraints, this corresponds
to a redefinition of the Lagrange multipliers in the total
Hamiltonian ${H}_T$
\eref{ht}. In the Hamiltonian PI \eref{hpi}
this is justified due to the presence of the $\delta$-functions.}
one finds
\be
\tilde{H}=H.
\label{e2}
\ee
Furthermore, the Lagrange
multipliers of the constraints \eref{sc}
with $i=1,\ldots,R$
that can be solved for $\tilde{Q}_{i,N+1}$,
i.e.\ that can be rewritten as
\be
\tilde{Q}_{i,N+1}-f_i(\tilde{Q}_{j,1},
\ldots ,\tilde{Q}_{j,N}, \tilde{P}_{j,N})=0
\qquad i=1,\ldots,R
\label{solve}
\ee
become zero in order to assure
$\dot{\tilde{P}}_{i,N+1}=0$
(because $\tilde{H}$ and the other constraints
do not depend on the $\tilde{Q}_{i,N+1}$ anymore).
The remaining of the constriants \eref{sc} are identical
to the constraints obtained in the $N$th order formalism.
Therefore, also the total Hamiltonians
which imply the EOM are identical%
\footnote{Actually, $\tilde{H}_T$ also contains Lagrange multipliers
corresponding to the primary constraints \eref{pc}. These however do
not affect the EOM \eref{qeom}, \eref{peom} for the
variables $\tilde{Q}_{i,n}$ and $\tilde{P}_{i,n}$
with  $n=1,\ldots,N$.}
\be
\tilde{H}_T=H_T.
\label{e3}
\ee
{}From \eref{e1}, \eref{e2} and \eref{e3} follows the equivalence of
the $N$th and the $(N+1)$st order Ostrogradsky formalism.

To extend this classical result to quantum physics
one has to show that the Hamiltonian PIs
obtained within the $N$th and the $(N+1)$st order
formalism are identical. In the
$(N+1)$st order formalism the Hamiltonian PI has the form
\bea
\tilde{Z}&\!\!=&\!\!
\int\prod_{i=1}^I\prod_{n=1}^{N+1}(\dfun \tilde{Q}_{i,n}
\dfun \tilde{P}_{i,n})
\exp\left\{i\int d^4x\,\left[-\tilde{H}+
\sum_{i=1}^I\sum_{n=1}^{N+1}(
\tilde{P}_{i,n}\dot{\tilde{Q}}_{i,n}+
J_{i,n}\tilde{Q}_{i,n}+K_{i,n}\tilde{P}_{i,n})\right]\right\}
\nn&&\mbox{\hspace*{7cm}}\times
\prod_a\delta( \tilde{\Phi}_a)
\,{\rm Det}^\frac{1}{2}\,\{\tilde{\Phi}_a,\tilde{\Phi}_b\}.
\label{hpi2}\eea
The constraints and g.f.\ conditions
$\tilde{\Phi}_a$ include all the constraints
and g.f.\ conditions $\Phi_a$
obtained in the $N$th
order formalism (because \eref{PN} and \eref{sc}
are identical) and, in addition, the constraints
\eref{pc} and \eref{solve}. The fundamental Poisson brackets imply
that \eref{pc} with $i=1,\ldots, R$ and \eref{solve} represent a
system of second class constraints, while the constraints \eref{pc}
with $i=R+1,\ldots,I$,
which have vanishing Poisson brackets with all
other constraints, are first class.
Therefore one has to include
gauge fixing conditions corresponding to these
first class constraints
into the $\tilde{\Phi}_a$. A possible and suitable choice is
\be
\tilde{Q}_{i,N+1}=0,\qquad i=R+1,\ldots,I.
\label{gf}
\ee
Using the fundamental poisson brackets and remembering that the
$\Phi_a$ do not depend on the $\tilde{Q}_{i,N+1}$
and the $\tilde{P}_{i,N+1}$ , one finds
\be
{\rm Det}\, \{\tilde{\Phi}_a,\tilde{\Phi}_b\}=
{\rm Det}\, \{\Phi_a,\Phi_b\}.
\ee
Rembering the presence of
the $\delta$-functions corresponding to the constraints
\eref{pc}, \eref{solve} and the g.f.\ conditions
\eref{gf} in the PI,
one can perform the functional
integrations over the $\tilde{Q}_{i,N+1}$ and the
$\tilde{P}_{i,N+1}$.
The equalities \eref{e1} and \eref{e2} imply that the
Hamiltonian PI \eref{hpi2} is identical to the one
obtained in the $N$th
order formalism\footnote{The sources $J_{i,N+1}$
with $i=1,\ldots,R$ are still present in $\tilde{Z}$ but this has no
physical effect.} \eref{hpi}.  Thus, {the
Hamiltonian PI is also independent of the formal order $N$.}

The conclusion of this section, which is most important for the
subsequent investigations is the following: In the $N$th order
formalism all derivatives up to the order $N-1$ are treated
as independent coordinates and
not as derivatives. Furthermore, the order $N$ can be
chosen arbitrarily
high without affecting the physical content of the
theory. This implies that {each local coordinate transformation
which also involves derivatives of the coordinates (up to a finite
order) can be considered to be a point transformation,
i.e. a transformation which formally only involves
the coordinates but not the derivatives.
If one wants to apply a coordinate
transformation involving derivatives up to the order $N$, one simply
has to treat the Lagrangian as an $(N+1)$st order one (even if
no $(N+1)$st order derivatives occur in $L$)
so that this transformation can be identified as
a point transformation.
Such a transformation
becomes a canonical transformation within the
Hamiltonian framework\/} because the fact that two Lagrangians
are related by a point transformation implies that the
corresponding Hamiltonians and constraints are related by a
canonical transformation\footnote{This can be most easily
seen if the order $N$ is choosen so high that the transformations
of the $Q_{i,n}$ with $n<N$ do not involve the $Q_{i,N}$.
\eref{ham} and \eref{PN} imply then that the Hamiltonian and the
primary constraints \eref{PN} with $i=R+1,\ldots,I$ are related
by a canonical transformation. The secondary, tertiary, etc.\
constraints follow
from the Poisson brackets of the primary constraints with the
Hamiltonian and each other
which are invariant under canonical transformations.}.
Since the physical content of a theory is not affected
by canonical transformations, one finds
that {\em Lagrangians, which are related by a local
coordinate transformation are physically
equivalent even if this
transformation involves  derivatives.}

The Hamiltonian PIs corresponding to Lagrangians
that are related by such a field transformation
are identical due to the invariance of the
Hamiltonian PI under canonical transformations \cite{gity,ham} and
its independence of the (formal) order $N$ (see above). Therefore,
{\em this equivalence is also valid in quantum physics.}


\section{Reduction of Higher Order Effective Lagrangians and
Matthews's Theorem}
\typeout{Section 3}
\eqnew
In this section I will show  that
a higher order effective Lagrangian can be reduced
to a first order one by applying the equations of
motion to the effective interaction term.
Using this reduction, I will generalize Matthews's theorem to
higher order effective Lagrangians. For simplicity I will first only
consider the case of a massive scalar field.

The effective Lagrangian has the form
\be
\lag=\lag_0+\ep\lag_I=\frac{1}{2}(\pa^\mu\vp)(\pa_\mu\vp)-
\frac{1}{2}M^2\vp^2+\ep\lag_I(\vp,\pa^\mu\vp,\,\ldots,\,
\pa^{\mu_1}\cdots\pa^{\mu_N}\vp).
\label{leff}
\ee
$\lag_0$ simply represents a free massive Klein--Gordon theory and
$\lag_I$ contains the effective interactions which depend on
the derivatives of the scalar fields up to the order $N$. These
interactions are governed
by the coupling constant $\ep$ with $\ep\ll 1$.

Now I want to remove all higher
order time derivatives from the effective
interaction term by applying the
EOM (upon neglecting higher powers of
$\ep$). This procedure must be carefully justified
since, in general, the EOM must not be used to convert
the Lagrangian. Therefore, I use the results of \cite{eom,pol,geo}
where it has been shown
that it is always possible to find
field transformations which effectively result
in applying the EOM following from
$\lag_0$ to $\lag_I$ (in the first order of $\epsilon$).
Assume that $\ep\lag_I$ contains a term
\be
\ep T\ddot{\vp}
\label{term}\ee
(where $T$ is an arbitrary expression in $\vp$ and its derivatives),
i.e.\ that \eref{leff} can be written as
\be
\lag=\lag_0+\ep T\ddot{\vp}+\ep\tilde{\lag}_I
\label{l1}
\ee
(with $\tilde{\lag}_I\equiv\lag_I-T\ddot{\vp}$).
Performing the field transformation \cite{eom,pol,geo}
\be
\vp\to\vp+\ep T
\label{trafo} \ee
one finds
\bea
\lag&=&\lag_0+\ep T\ddot{\vp}+\ep T\left(\frac{\pa\lag_0}{\pa\vp}-
\pa^\mu\frac{\pa\lag_0}{\pa(\pa^\mu\vp)}\right)+
\ep\tilde{\lag}_I+O(\ep^2)\nonumber\\
&=&\lag_0+\ep T (\Delta \vp-M^2 \vp)
+\ep\tilde{\lag}_I+O(\ep^2).
\eea
This means
that effectively the second order time derivative has been removed
from the term \eref{term} by applying
the free EOM (i.e.\, those
implied by $\lag_0$ alone)
\be
\ddot{\vp}=\Delta\vp-M^2\vp.
\label{kgeom}\ee
If there are terms with
higher than second order derivatives in $\ep\lag_I$,
they can be put into the form
\eref{term} by
performing partial integrations and dropping total derivative
terms\footnote{In general, only
total derivatives of expressions that
depend on nothing but the coordinates can be dropped.
However, since the
derivatives are treated as coordinates
within the Ostrogradsky formalism
if the order $N$ is chosen sufficiently high (as discussed in the
previous section),
all total derivative terms can be neglected
\cite{gity}.}.
Repeating the above procedure, one can remove all higher order
time derivatives from the effective Lagrangian.

To prove that a higher order effective Lagrangian and
the first order Lagrangian obtained from it by applying the
above procedure are physically equivalent, I
apply the results of the previous section. The Lagrangians are
connected by field transformations of the type \eref{trafo} which,
in general, involve derivatives of $\vp$ (contained in $T$).
However, within
the Ostrogradsky formalism the transformations are point (i.e.\
canonical) transformations (if the formal order of $\lag$ is
choosen sufficiently high)
which establishes the equivalence of both
Lagrangians. To be strictly correct, one must remember that, within
the Ostrogradsky formalism, the time derivatives of $\vp$ are
fields that are formally independent  of $\vp$.
Therefore, the transformation of
$\dot{\vp}$ must be specified seperately:
\be
\dot{\vp}\to\dot{\vp}+\ep \dot{T},
\label{trafo2}\ee
because this relation is not  automatically implied by \eref{trafo}
as in the first order formalism. \eref{trafo} and \eref{trafo2}
specify the field transformation completely (since,
due to the neglection of higher powes of $\ep$, this transformation
is de facto only appplied to $\lag_0$
which contains at most first order
derivatives). Now one can
easily see that, if $T$ contains at most $M$th order derivatives,
the Lagrangian formally has to be treated as an at least
$(M+2)$nd order one so that \eref{trafo}, \eref{trafo2} becomes
a canonical transformation.
The result of the above procedure is that that
{\em the equations of motion
following from $\lag_0$ may be applied to convert
$\lag_I$ in order
to eliminate all higher order time
derivatives} (upon neglecting higher powers of
$\ep$).

Now it is easy to quantize the effective Lagrangian \eref{leff}
and to prove Matthews's theorem \eref{lpi} with \eref{eq}.
The proof goes as
follows:
\begin{enumerate}
\item Given an effective higher order Lagrangian $\lag$ \eref{leff},
it can be reduced by
applying the above procedure to an equivalent
first order\footnote{Formally, the reduced Lagrangian $\lag_{red}$
still has to be treated as
an $N$th order one although there are no
more higher order derivatives. But, as shown in
section~2, it does not affect the physical content
of the theory to treat it as
a first order Lagrangian.} Lagrangian $\lag_{red}$.
As discussed in section~2 this does not affect the Hamiltonian PI.
\item $\lag_{red}$ can be
quantized by applying Matthews's theorem for
first order Lagrangians \cite{bedu}. The generating functional%
\footnote{Since $\lag_{red}$
is a first order Lagrangian, source terms only have to be introduced
for the field $\vp$ but not for its derivatives as in the general
PI formalism for higher order Lagrangians.}
can be written as \eref{lpi} with the quantized Lagrangian
\be
\lag_{quant}=\lag_{red}.
\ee
\item Within the PI, the field transformations \eref{trafo} become
performed inversely in order to reconstruct the primordial higher
order Lagrangian. This finally yields \eref{eq}.
\end{enumerate}
The last step needs some additional clarification because a field
transformation like \eref{trafo}, if it is performed within the
Lagrangian PI
\eref{lpi}, does not only affect the quantized
Lagrangian, but also the integration measure and the source terms.
However, the transformation of the measure
only yields extra $\delta^4(0)$
terms \cite{pol} which are neglected here (see introduction and
\cite{bedu,gk}) and the change in the source terms does
not affect the physical matrix elements \cite{pol,able}.

Matthews's theorem implies, that an {\em effective\/} higher order
Lagrangian can be quantized in the same way as
a first order one without worrying about the
unphysical effects that are
normally connected with higher order Lagrangians.
In particular, the
Feynman rules can be obtained
from the effective interaction terms in
the standard manner.

Closing this section, I want to add some remarks:
\begin{itemize}
\item This formalism can only be applied if higher powers of
$\epsilon$ are neglected, since it is possible to eliminate the
higher order time derivatives in the first order of $\ep$
(and in fact in any finite order of $\ep$ \cite{geo}) but they
cannot
be removed completely. As mentioned in the introduction, this
treatment is justified within the effective Lagrangain formalism
because effects implied by $O(\epsilon^n)$ terms with $n>1$
are assumed to be
cancelled by other effects of (well-behaved) ``new physics''.
\item The Ostrogradsky
formalism itsself is in principle a reduction of
a higher order Lagrangian to a first order one because the higher
order derivatives are considered to be independent coordinates;
however, this means that new degrees of
freedom are introduced. These additional degrees of
freedom involve the unphysical effects \cite{hide,bedu}.
Here, an
effective higher order Lagrangian is reduced to a first order one
{\em without\/} introducing extra degrees of freedom.
\item The use of the EOM \eref{kgeom} may, in general, yield
expressions  in $\lag_{red}$
which are not manifestly covariant (see especially the
treatment of the following section).
Also the Hamiltonian PI quantization
procedure involves such
terms (see \cite{bedu,gity,ham,gk}). However, these
expressions only occur in intermediate steps
of the derivation but not
in the resulting expression \eref{lpi} with \eref{eq}.
Actually, Matthews theorem enables calculations based
on the manifestly covariant Lagrangian PI.
\end{itemize}


\section{Higher Order Effective Interactions of
Massive Vector Fields}
\typeout{Section 4}
\eqnew
In this section I will generalize the results of the preceding
one to higher order effective (non Yang--Mills)
self-interactions of massive vector fields. I will examine
the three different types of effective Lagrangians which are studied
in
the literature, namely gauge noninvariant Lagrangians, SBGTs with a
nonlinearly realized scalar sector (gauged nonlinear
$\sigma$-models) and SBGTs with a linearly realized scalar sector
(Higgs models).

For simplicity, I will only consider
massive Yang--Mills fields (where all vector bosons have equal
masses) with additional effective (non--Yang--Mills)
interactions and the corresponding SBGTs, as in \cite{gk}.
The generalization to
other, e.g. electroweak, models is straightforward.


\subsection{Gauge Noninvariant Models}
I consider the effective Lagrangian
\be
\lag=\lag_0+\ep\lag_I=-\frac{1}{4}
F^{\mu\nu}_aF_{\mu\nu}^a+\frac{1}{2}M^2A_a^\mu
A_\mu^a+\ep\lag_I (A_a^\mu,\pa^\nu A_a^\mu,
\, \ldots\, , \pa^{\nu_1}\cdots\pa^{\nu_N}A_a^\mu).
\label{ymeff}
\ee
The field strength tensor is given by
\be
F^{\mu\nu}_a\equiv\pa^\mu A_a^\nu-\pa^\nu A_a^\mu
+gf_{abc}A^\mu_b A^\nu_c.
\ee
$\lag_0$ represents a massive Yang--Mills theory and  the effective
interaction term $\lag_I$ contains the deviations from the
Yang--Mills interactions which involve derivatives up to the order
$N$ and which are proportional to $\ep$ with $\ep\ll 1$.

By applying the procedure described in section~3 one
can now use the EOM following from
$\lag_0$  to eliminate the higher order
time derivatives in
$\lag_I$. These
EOM are:
\be
D_\mu F^{\mu\nu}_a=-M^2A_a^\nu
\label{ymeom}
\ee
with the covariant derivative
\be
D^\sigma F^{\mu\nu}_a\equiv \pa^\sigma F^{\mu\nu}_a
+gf_{abc}A^\sigma_b F^{\mu\nu}_c.
\label{df}\ee
With $\nu=i=1,2,3$, \eref{ymeom} can be rewritten as
\be
\ddot{A}^i_a=-M^2A^i_a-D_jF^{ij}_a+gf_{abc}A_b^0F^{i0}_c
-\pa_i\dot{A}_a^0+gf_{abc}(\dot{A}_b^iA_c^0
+A^i_b\dot{A}_c^0).
\label{ddai}
\ee
This equation serves to eliminate all higher order
time derivatives of the
$A^i_a$. Now one has to get rid of the time derivatives of the
$A^0_a$.
To be able to apply Matthews's theorem for Lagrangians without
higher order time derivatives \cite{gk}, one even has to remove the
first order time derivatives
of the $A_a^0$ because in \cite{gk} it is
assumed that $\lag$ does not depend on $\dot{A}_a^0$ (see
footnote~3 there).
With $\nu=0$, \eref{ymeom} becomes
\be
A_a^0=\frac{1}{M}[-\pa_i F^{i0}_a+gf_{abc}A_b^i F^{i0}_c].
\label{a0}
\ee
differentiation yields
\be
\dot{A}_a^0=\frac{1}{M}[{}-\pa_i\dot{F}^{i0}_a+
gf_{abc}(\dot{A}_b^i F^{i0}_c+A_b^i \dot{F}^{i0}_c)],
\label{da0}
\ee
where $\dot{F}^{i0}_a$ can be written, using \eref{ddai}, as
\be
\dot{F}^{i0}_a=M^2 A^i_a+ D_jF^{ij} - gf_{abc} A^0_b F^{i0}_c.
\label{dfi0}\ee
By repeated application of \eref{ddai}, \eref{da0} and \eref{dfi0}
one can reduce the effective Lagrangian \eref{ymeff} to an
equivalent Lagrangian $\lag_{red}$ which contains neither higher
order time derivatives of the vector fields
nor first order time derivatives of the $A_a^0$.

Now Matthews's theorem can be proven as in the previous section. The
effective higher order Lagrangian becomes reduced to a first order
Lagrangian $\lag_{red}$ as described above,
this gets quantized using Matthews's
theorem for effective first order interactions of massive vector
fields \cite{gk} and finally one can reconstruct
the primordial Lagrangian $\lag$
by applying appropriate field transformations within
the Lagrangian PI.


\subsection{Gauged Nonlinear $\bf \sigma$-Models}
Now I consider SBGTs
with additional effective
interaction terms  that are embedded into a gauge
invariant framework. First I restrict to models
that do not contain physical Higgs fields, which implies that
the unphysical pseudo-Goldstone fields $\vp_a$
are nonlinearly realized as
\be
U=\exp\left(i\frac{g}{M}\vp_a t_a\right),
\label{U}\ee
where the $t_a$ are the generators of the gauge group. Within these
models, it is most convenient to use the matrix notation defined by:
\be
A^\mu\equiv A^\mu_a t_a,\qquad F^{\mu\nu}\equiv
F^{\mu\nu}_a t_a,
\qquad\mbox{etc.}
\ee

In \cite{bulo,gkko}, it has been
shown that each effective Lagrangian
of the type \eref{ymeff} can be rewritten as a nonlinear SBGT by
applying the Stueckelberg transformation \cite{stue}
\be
A^\mu\to-\frac{i}{g}U^\dagger D^\mu U
\label{stue}\ee
with the covariant derivative
\be
D^\mu U\equiv \pa^\mu U+ig A^\mu U.
\label{du}
\ee
On the other hand, each effective gauged
nonlinear $\sigma$-model can be
rewritten in the gauge noninvariant form
\eref{ymeff} by inversing the Stueckelberg
transformation \eref{stue}.
This can be seen as follows: due to the gauge invariance
of the effective interaction term, the fields only occur there
in the gauge invariant combinations
\bea &&
U^\dagger(D^{\sigma_1}\cdots D^{\sigma_N}F^{\mu\nu})U,
\label{ddF}\\&&
U^\dagger D^{\sigma_1}\cdots D^{\sigma_N}U.
\label{ddu}\eea
(The higher order covariant derivatives of $F^{\mu\nu}$ and $U$
are defined analogously to the first order ones \eref{df} and
\eref{du}.) Each effective interaction term can be constructed
from the expressions \eref{ddF},
\eref{ddu} and constants like $t_a$, $g^{\mu\nu}$ and
$\ep^{\rho\sigma\mu\nu}$
by taking products, sums, derivatives and traces \cite{apbe}.
The term \eref{ddF} becomes
\be
D^{\sigma_1}\cdots D^{\sigma_N}F^{\mu\nu}
\ee
after an inverse Stueckelberg transformation.
By using the unitarity of $U$ \eref{U},
\be
UU^\dagger=1,
\ee
and performing product differentiation, the term \eref{ddu} can be
expressed through terms
\be
U^\dagger D^\sigma U \label{DU}
\ee
and their derivatives.
E.g.\ for $N=2$ it can be written as
\be
(U^\dagger D^{\sigma_1} U)(U^\dagger D^{\sigma_2} U)
+\pa^{\sigma_1}(U^\dagger D^{\sigma_2} U).
\ee
(Similar formulas can be found for $N>2$).
\eref{DU} becomes
\be
ig A^\sigma
\ee
after an inverse Stueckelberg transformation.
This means that each nonlinear gauge invariant Lagrangian
$\lag^S$ can be rewritten as
a gauge noninvariant Lagrangian $\lag_U$
(U-gauge Lagrangian)
by applying the inverse of the
Stueckelberg transformation \eref{stue}. As
one can see from the above procedure, this transformation
results in simply dropping all unphysical scalar fields in
$\lag^S$:
\be
\lag_U\equiv\lag^S\Big |_{A^\mu\to U A^\mu U^\dagger
-\frac{i}{g}U\pa^\mu U^\dagger}
=\lag^S\Big|_{\vp_a=0}.
\label{ug}
\ee

In \cite{gk}, I have proven that Lagrangians
(without higher order derivatives)
which are related by a Stueckelberg transformation
are equivalent within the Hamiltonian formalism.
Within the Ostrogradsky formalism
this result can be generalized to higher order Lagrangians
and besides it can be derived more easily.
Since a Stueckelberg transformation
is a field transformation that depends on the derivatives of the
fields, it is a  canonical
transformation within this formalism\footnote{One may wonder
why a gauge invariant (i.e.\ first class constrained) system
can be related to a gauge noninvariant (i.e.\
second class constrained)
system by a canonical transformation because the number of second
class constraints is given by ${\rm rank}\,\{\phi_a,\phi_b\}|_
{\scriptstyle \phi_a=0}$ which is invariant under canonical
transformations. One should remember that $\lag^S$ and $\lag_U$
are only related by a canonical transformation if the order $N$ is
artificially increased. This procedure yields additional constraints
(see section~2). In fact, $\lag^S$ and $\lag_U$ imply equal numbers
of first class and of second class constraints if $N$ is choosen
sufficiently high.}
(if the formal
order of $\lag^S$
is chosen sufficiently high). The transformations of the time
derivatives of $A^\mu$ must be specified seperately because
these
are considered to be independent fields. With $M$ being the
order of the
highest time derivative of $A^\mu $ appearing in $\lag^S$, the
inverse
Stueckelberg transformation is then completely specified by
\be
\pa^m_0A^\mu
\to\pa^m_0\left(U A^\mu U^\dagger
-\frac{i}{g}U\pa^\mu U^\dagger\right)
,\qquad m=0,\ldots,M.
\label{stue2}
\ee
The Lagrangian $\lag^S$ has thus formally to be treated as an
at least $(M+2)$nd order one to establish the equivalence of
$\lag^S$ and $\lag_U$ in \eref{ug}.

Matthews's theorem for effective
gauged nonlinear $\sigma$-models with higher
order derivatives can now be proven as follows:
Given a nonlinear gauge invariant Lagrangian $\lag^S$, this can be
converted into an equivalent gauge noninvariant Lagrangian $\lag_U$
\eref{ug} by using the Stueckelberg formalism.
(As discussed in section~2, this does
not affect the Hamiltonian path integral.)
Since $\lag_U$ is of the type
\eref{ymeff}, the results of the previous subsection can be applied
to quantize it. This yields
the generating functional \eref{lpi} with
\be
\lag_{quant}=\lag_U.
\label{ugq}
\ee
This, however, is identical to
the result of the Faddeev--Popov quantization
procedure applied to $\lag^S$ if the (U-gauge) g.f.\
conditions
\be
\vp_a=0
\label{ugf}\ee
are imposed \cite{gkko}. Due to the equivalence of all gauges
\cite{lezj,able}, this result
can be rewritten in any other gauge%
\footnote{Actually, loop calculations within
the U-gauge suffer from ambiguities in the determination of the
finite part of an $S$-matrix element \cite{jawe,fls}. Therefore, for
practical calculations it is useful to rewrite the PI obtained
within in the
U-gauge in the $\rm R_\xi$-gauged form
in order to remove these
ambiguities. In fact, loop calculations within the U-gauge yield the
same $S$-matrix elements as loop calculations
within the $\rm R_\xi$-gauge, but only if the
correct renormalization prescription is used
\cite{fls}; other renormalization prescriptions yield
distinct results.
In the $\rm R_\xi$-gauge these ambiguities are
absent.}
(e.g.\
$\rm R_\xi$-gauge, Lorentz gauge, Coulomb gauge, etc.).
This completes the proof of Matthews's theorem for effective gauged
nonlinear $\sigma$-models.


\subsection{Higgs Models}
Finally let me consider effective
SBGTs with linearly realized symmetry, i.e.\ models that
involve additional physical scalar fields
and that also contain gauge
invariant effective interaction terms. Since these models cannot be
written in a general form for an arbitray gauge group, I restrict to
the case of SU(2) symmetry (i.e.\ $t_a=\frac{1}{2}\tau_a$,
$a=1,2,3$) as in \cite{gk}.
The generalization to another gauge group
is straightforward.

Within an SU(2) Higgs model the scalar sector is parametrized as
\be
\Phi=\frac{1}{\sqrt{2}}((v+h){\bf 1}+i\vp_a\tau_a)
\label{higgslin}\ee
with the Higgs field $h$ and the vacuum expectation value
$v=\frac{2M}{g}$. However, even within a linear Higgs model the
scalar sector can be parametrized nonlinearly by performing the
transformation \cite{gk,gkko,lezj,clt}
\be
\Phi\to\frac{v+h}{\sqrt{2}}U=\frac{v+h}
{\sqrt{2}}\exp\left(\frac{i
\vp_a\tau_a}{v}\right).
\label{higgsnonlin}\ee
Applying this transformation to the Lagrangian $\lag^H$
of an effective Higgs model, this
becomes converted into an effective nonlinear Stueckelberg model
(as examined in the previous subsection)
which contains an additional
physical scalar field. Lagrangians which are related by the
transformation
\eref{higgsnonlin} are clearly equivalent because \eref{higgsnonlin}
is a point (i.e.\ canonical)
transformation. By applying a Stueckelberg transformation,
the resulting nonlinear Lagrangian
can be reduced to an equivalent gauge nonivariant
Lagrangian $\lag_U$ (U-gauge Lagrangian)
which is obtained by dropping all unphysical
scalar fields in $\lag^H$:
\be
\lag_U\equiv
\lag^H\Bigg|_{\ba{l}\scriptstyle \Phi\to\frac{v+h}
{\sqrt{2}}U \\
\scriptstyle\vp_a=0\ea}=\lag^H\Big|{\scriptstyle\vp_a=0}.
\label{uh}\ee
$\lag_U$ is similar to \eref{ymeff} but it
contains an additional physical scalar field $h$. Therefore
$\lag_U$ is of the form
\bea
\lag_U&=&\lag_0+\ep\lag_I\nn&=&-\frac{1}{4}
F^{\mu\nu}_aF_{\mu\nu}^a+\frac{1}{2}(\pa^\mu h)(\pa_\mu h)
+\frac{1}{8}g^2 (v+h)^2A_a^\mu
A_\mu^a
\nn&&{}-\frac{1}{2}M_H^2 h^2
-\frac{1}{4}g\frac{M_H^2}{M}h^3-\frac{1}{32}g^2
\frac{M^2_H}{M^2}h^4\nn&&{}
+\ep\lag_I (A_a^\mu,\pa^\nu A_a^\mu,
\, \ldots\, , \pa^{\nu_1}\cdots\pa^{\nu_N}A_a^\mu
,h,\pa^\mu h,\,\ldots\, ,\pa^{\mu_1}\cdots\pa^{\mu_N}h).
\eea
$\lag_0$ is the U-gauge Lagrangian of the Higgs model without
effective interaction terms.
The EOM corresponding to \eref{ddai}, \eref{da0}, \eref{dfi0}
and the EOM for $h$ that follow from $\lag_0$ are
\bea
\!\!\!\!\!\!\!\!\!\!\!\!\!
\ddot{A}^i_a&=&{}-\frac{1}{4}g^2(v+h)^2
A^i_a-D_jF^{ij}_a+gf_{abc}A_b^0F^{i0}_c
-\pa_i\dot{A}_a^0+gf_{abc}(\dot{A}_b^iA_c^0
+A^i_b\dot{A}_c^0),\\\!\!\!\!\!\!\!\!\!\!\!\!\!
\dot{A}_a^0&=&\frac{4}{g^2(v+h)^2}\left[-\frac{1}{2}
g^2(v+h)\dot{h}A^0_a-\pa_i\dot{F}^{i0}_a+
gf_{abc}(\dot{A}_b^i F^{i0}_c+A_b^i
\dot{F}^{i0}_c)\right],\label{nonpol}\\\!\!\!\!
\!\!\!\!\!\!\!\!\!
\dot{F}^{i0}_a&=& \frac{1}{4}g^2(v+h)^2
A^i_a+ D_jF^{ij} - gf_{abc} A^0_b F^{i0}_c,
\\\!\!\!\!\!\!\!\!\!\!\!\!\!
\ddot{h}&=&\Delta h
+\frac{1}{4}g^2(v+h)A_a^\mu A^a_\mu-M_H^2 h
-\frac{3}{4}g\frac{M_H^2}{M}h^2-\frac{1}{8}g^2\frac{M^2_H}{M^2}h^3.
\eea
These EOM have to be used to eliminate all higher order time
derivatives of the $A^\mu_a$ and of $h$
and also the first order time derivatives of the $A^0_a$ in
$\lag_U$ \eref{uh}.

$\lag^H$ can now be quantized by using the procedure
described in the
previous subsections and the results of
\cite{gk}. One obtains
the Lagrangian PI \eref{lpi} with
\be
\lag_{quant}=\lag_U.
\ee
This is again identical to the result of the
Faddev--Popov formalism
with the (U-gauge) g.f.\ conditions \eref{ugf}
\cite{gkko} (upon neglecting a $\delta^4(0)$ term).
Using the equivalence of all gauges \cite{lezj,able},
this result can be generalized to any other gauge.
Thus, the proof of
Matthews's theorem is extended to Higgs models with
effective higher order interactions.

It should be mentioned
that the transformation \eref{higgsnonlin} and
the EOM \eref{nonpol}
involve nonpolynomial interactions, which are not
present in the primordial Lagrangian. This, however, is no serious
problem, since these expressions only
occur in intermediate steps of the derivation
but not in the resulting Faddeev-Popov PI
\eref{lpi} with \eref{fapo}. The
application of Matthews's theorem for first order Lagrangians is not
affected by these terms, since the treatment of \cite{gk} does not
necessarily require polynomial interactions.

The procedure of this subsection illustrates that the
above proof can be extended to Lagrangians that also contain
couplings to matter fields (which have been negelcted here for
simplicity). The additional
couplings in $\lag_0$ affect the EOM, but
the general statement
that each effective higher order Lagrangian can
be reduced to a first order one by using the EOM remains unaffected.
Therefore, {\em for any effective Lagrangian, Hamiltonian
and Lagrangian path integral quantization are equivalent.}


\section{Conclusions}
\typeout{Section 5}
\eqnew
In this paper I have shown that the unphysical effects
connected with Lagrangians that depend on higher order derivatives
are absent if an {\em effective\/} Lagrangian  is considered,
i.e. a Lagrangian that represents the
low energy approximation of well-behaved ``new physics''
with heavy particles at a higher energy scale.

Upon neglecting higher
powers of the effective coupling constant $\epsilon$,
all higher order time derivatives can be eliminated
from an effective Lagrangian. This reduction of a
higher order effective
Lagrangian to a first order one is done by applying
of the equations of motion to the effective interaction term.
The (in general forbidden) use of the
equations of motion is justified within the effective Lagrangian
formalism because
it can be realized by performing field transformations
that involve derivatives of the fields.
Lagrangians that are
related by such a field transformation are physically
equivalent (at the classical and at the quantum level)
because these transformations
are canonical transformations within the Hamiltonian treatment
of higher order Lagrangians (Ostrogradsky formalism).

Matthews's theorem has been extended to higher order effective
Lagrangians by applying this reduction and
using Matthews's
theorem for first order effective Lagrangians.  This theorem states
that any effective Lagrangian can be quantized using the
simple Lagrangian path intergral ansatz \eref{lpi} with \eref{eq}
(or the Faddeev--Popov
formalism, \eref{lpi} with \eref{fapo}, in the
case of gauge invariant effective Lagrangians), since this  turns
out to be the result of the correct Hamiltonian path integral
quantization procedure. The Feynman rules can be
directly obtained
from the effective Lagrangian in the standard manner.

Thus, {\em effective\/} higher
order Lagrangians can be treated in the
same way as first order
Lagrangians, the higher order terms do not imply
any unphysical effects
unlike in the general treatment of higher order Lagrangians.

Finally, two important points should be noted:
\begin{itemize}
\item {\em
The formalism described in this paper can only be applied
to effective Lagrangians\/}, since then the supposed existence
of well-behaved ``new physics'' beyond the theory described by the
effective Lagrangian
justifies the omission of all
unphysical effects. In fact, the assumption $\ep\ll 1$
{\em alone} is not
sufficient for neglecting higher powers of $\epsilon$ since
theories with higher order derivatives have no analytic limit
for $\ep\to 0$. Thus, the effects of a term with higher order
derivatives are not small even if the coupling constant of this
term is extremely small \cite{hide,bedu}. This implies that
the unphysical
effects cannot be avoided within models with higher order
derivatives that are not considered to be effective ones.
\item Since each effective Lagrangian with higher order derivatives
can be reduced to a first order Lagrangian, it is in
principle sufficient to consider only effective Lagrangians with
at most first order derivatives \cite{geo,buwy,ruj}. However,
the reduction of a quite simple higher order effective
interaction term to a first order term
by applying the equations of motion,
in general,
yields a lengthy and awkward expression.
Even more, in the reduced form the physical effects of such a term
are not very obvious because in this form it becomes
a linear combination of several terms; each one alone of these terms
yields effects that are not implied by the primordial term
but among them complicated
cancellations take place \cite{ruj,gkks}.
Thus, for practical purposes
it is much more convenient to work with the primordial higher order
Lagrangian instead of the reduced first order Lagrangian. Therefore,
I have used this reduction only
for technical purposes in order to apply
Matthews's theorem for
first order Lagrangians, but in the final result
\eref{lpi} with \eref{eq} or \eref{fapo} I have reconstructed
the original higher order Lagrangian. Actually, this result
enables calculations based on a higher order effective
Lagrangian without doing this reduction.
\end{itemize}

\section*{Acknowledgement}
I like to
thank Reinhart  K\"ogerler for many
helpful discussions and for reading
the manuscript of this paper.




\begin{thebibliography}{00}
\typeout{References}
\bibitem{apbe}T. Appelquist and C. Bernard,
\phrd{22} (1980) 200;\\
A. C. Longhitano, \nphb{188} (1981) 118
\bibitem{phen}C. N. Leung, S. T. Love and S. Rao,
\zphc{31} (1986) 433;\\
B. Grinstein and M. B. Wise, \phlb{265} (1991) 326;\\
G. J. Gounaris and F. M. Renard, Montpellier
Preprint PM/92-31 (1992);\\
K. Hagiwara, S. Ishihara, R. Szalapski and D. Zeppenfeld,
Madison Preprint MAD/PH/737 (1993)
\bibitem{ost}M. V. Ostrogradsky, Mem.\ Acad.\ St.\ Petersbourg {\bf
6} (1850) 385
\bibitem{hide}A. Pais and G. Uhlenbeck, \phr{79} (1950) 145;\\
S. W. Hawking, in ``Quantum Field Theory and Quantum
Statistics'', ed.: I. A. Batalin, C. J. Isham and C. A. Vilkovisky
(Hilger, 1987) p.~129;\\
D. A. Eliezer and R. P. Woodard, \nphb{325} (1989) 389;\\
J. Z. Simon, \phrd{41} (1990) 3720
\bibitem{bedu}C. Bernard and A. Duncan, \phrd{11} (1975) 848
\bibitem{eom}D. Barua and S. N. Gupta, \phrd{16} (1977) 413;\\
G. Sch\"afer, Phys.\ Lett.\ {\bf A100} (1984) 128;\\
T. Damour and G. Sch\"afer, J.\ Math.\ Phys.\ {\bf 32} (1991) 127;\\
H. Leutwyler, Bern Preprint BUTP-91/26 (1991)
\bibitem{pol}H. D. Politzer, \nphb{172} (1980) 349;\\
C. Arzt, Michigan-Preprint UM-TH-92-28 (1992),
hep-ph/9304230
\bibitem{geo}H. Georgi, \nphb{361} (1991) 339
\bibitem{gity}D. M. Gitman and I. V. Tyutin,
``Quantization of Fields with
Constraints'' (Springer, 1990)
\bibitem{ham}L. D. Faddeev, Teor.\ Mat.\ Fiz. {\bf 1}
(1969) 3 [Transl.:
Theor.\ Math.\ Phys.\ {\bf 1} (1970) 1];\\
P. Senjanovic, \aph{100} (1976) 227
\bibitem{gk}C. Grosse-Knetter, Bielefeld Preprint BI-TP 93/17 (1993),
hep-ph/9304310
\bibitem{mat}P. T. Matthews, \phr{76} (1949) 684
\bibitem{bnn}J. Barcelos-Neto and C. P. Natividade, \zphc{51}
(1991) 313
\bibitem{stue}E. C. G. Stueckelberg,
Helv.\ Phys.\ Acta {\bf 11} (1938) 299;\\
T. Kunimasa and T. Goto, \ptph{37} (1967) 452;\\
T. Sonoda and S. Y. Tsai, \ptph{71} (1984) 878
\bibitem{bulo}M. Chanowitz, M. Golden and H. Georgi,
\phrd{36} (1987) 1490 ;\\
C. P. Burgess and D. London, McGill Preprint McGill-92/04 (1992),
hep-ph/9203215; \phrl{69} (1992) 3428
\bibitem{gkko}C. Grosse-Knetter and R. K\"ogerler,
Bielefeld Preprint BI-TP
92/56 (1992), hep-ph/9212268
\bibitem{lezj}B. W. Lee and J. Zinn-Justin,
\phrd{5} (1972) 3121, 3137, 3155,
{\bf D7} (1973) 1049
\bibitem{clt}J. M. Cornwall, D. N. Levin and G. Tiktopoulos,
\phrd{10} (1974) 1145
\bibitem{fapo}L. D. Faddeev and V. N. Popov, \phlb{25} (1967) 29
\bibitem{able}E. S. Abers and B. W. Lee, \phrp{9} (1973) 1
\bibitem{shd}Y. Saito, R. Sugano, T. Ohta and T. Kimura,
J. Math.\ Phys.\ {\bf 30} (1989) 1122;\\
J. M. Pons, Lett.\ Math.\ Phys.\ {\bf 17} (1989) 181;\\
V. V. Nestorenko, J. Phys.\ {\bf A22} (1989) 1673
\bibitem{dirac}P. A.M. Dirac, Can.\ J. Math.\ {\bf 2} (1950) 129;
``Lectures on Quantum Mechanics'' (Belfar, 1964)
\bibitem{jawe}R. Jackiw and S. Weinberg, Phys.\ Rev.\ {\bf D5}
(1972) 2396;\\
I. Bars and M. Yoshimura, Phys.\ Rev.\ {\bf D6} (1972) 374
\bibitem{fls}K. Fujikawa, B. W. Lee and A. I. Sanda,
Phys.\ Rev.\ {\bf D6} (1972) 2923
\bibitem{buwy}W. Buchm\"uller and D. Wyler, \nphb{268} (1986) 621
\bibitem{ruj}A. de R\'{u}jula,
M. B. Gavela, P. Hern\'{a}ndez and E. Mass\'{o},
\nphb{384} (1992) 3
\bibitem{gkks}C. Grosse-Knetter, I. Kuss and D. Schildknecht,
Bielefeld Preprint BI-TP
93/15 (1993), hep-ph/9304281
\end{thebibliography}
\end{document}